\documentclass[twocolumn,showpacs,preprintnumbers,amsmath,amssymb]{revtex4}
\usepackage{color}
\usepackage{bm, graphicx, amsmath}
\usepackage{hyperref}
\usepackage{soul}

\begin{document}

\title{Detection of patterns in a discrete-outcome sensor network}
\author{Pranjal Agarwal$^{1,2}$, Nada Ali$^{1,2}$ and Mark Hillery$^{1,2}$}
\affiliation{$^{1}$Department of Physics, Hunter College of the City University of New York, 695 Park Avenue, New York, NY 10065 USA \\ $^{2}$Physics Program, Graduate Center of the City University of New York, 365 Fifth Avenue, New York, NY 10016}

\begin{abstract}
A discrete outcome quantum sensor network is one in which we are only interested in which detectors are activated.  This can be studied in either the strong or weak interaction regime.  If the detectors interact strongly with the environment, it is possible to definitely find which ones were activated.  If the interaction is weaker, there is a possibility of making an error, and the object is to minimize the probability of this happening.  Here we will be interested in this weaker interaction regime.  We will also assume that only certain patterns of detectors will be activated, different patterns being translated versions of a fundamental one. Our object will be to find which pattern has been activated.  We will look at both one and two-dimensional detector arrays and make use of techniques from minimum-error state discrimination.
\end{abstract}

\maketitle

\section{Introduction}

There has been considerable recent work on quantum sensor networks, determining if, and when, quantum effects such as entanglement can provide an advantage over what can be accomplished with a classical network.  It was known that quantum effects, in particular nonclassical states of light, can be used to improve the performance of individual detectors, with the most spectacular example being the use of squeezed states to improve the sensitivity of the LIGO gravitational wave detector.  Initially the work on quantum sensor networks studied the following problem.  The detectors, which are quantum systems, have interacted with the environment, and, as a result, parameters due to this interaction are encoded in their state.  In almost all cases, these parameters have been taken to be continuous variables.  For example, they could be the strength of a magnetic field at different locations.  The object is to estimate the parameters or some function of them.  The detectors can be of various types, qubits \cite{gorshkov,qian}, continuous variable systems \cite{shapiro}, or general quantum systems \cite{proctor}.  It was found that for finite-dimensional systems, entanglement of the quantum systems does not provide an advantage in estimating the individual parameters, but does provide an advantage in estimating a function of them \cite{gorshkov,qian,proctor,Rubio}.  It has also been shown that entangled states in optical networks can provide an advantage for distributed sensing \cite{shapiro,Zhang}.  Further studies have investigated whether linear optical networks with unentangled inputs can give a quantum advantage in distributed metrology \cite{ge}, and whether continuous-variable error correction can be useful in protecting a network of continuous-variable sensors from the effects of noise \cite{preskill}.

More recently, a different problem was studied.  Suppose that instead of determining a parameter, one is interested in whether or not a detector has detected something or which detector has detected something.  This kind of problem is described by discrete rather than continuous variables, and is a problem in channel discrimination \cite{kitaev,acin,DAriano,sacchi1,sacchi2,wang,pirandola}.  Each detector in a network can receive an input or no input.  Suppose the unitary operator $U$ describes the interaction between an input and a detector.  The operator $U$ could describe, for example, the rotation of a spin caused by a magnetic field, or a phase shift induced in a state of light by a transparent object.  The different output states of the detectors will be produced by starting with the initial state of the detectors and applying to it the operator consisting of $U$ acting on one or more of the detectors and the identity acting on the rest.  We then want to measure the output state in order to determine which detector received an input.  This means that we have to optimize over both the initial state of the detectors, and the final measurement.   In the case that only one detector is excited, this problem was studied in \cite{hillery,zhan}.  What was found was that for a small number of detectors starting from an initially entangled state helps, but the advantage decreases with the number of detectors.  This was extended in two ways in \cite{ali}.  First, the case of more than one detector being excited was considered.  Second, it was shown that if the detectors are arranged into groups, and one is only interested in which group detectors are excited, entanglement confers an advantage.  Finally, a related problem of picking out a target quantum channel from a background of identical channels has been analyzed by Zhuang and Pirandola, and useful bounds on channel discrimination have been derived \cite{Zhuang1,Zhuang2,Pereira}.

In our analysis of quantum sensor networks we made use of minimum-error state discrimination to determine the output state of the network\cite{helstrom,review}.  In \cite{zhan} we noted that if the interaction strength between the detector and the environment is sufficiently large, the output states can be discriminated without error.  This was noticed independently in \cite{chuang1}.  They used it to define the single-shot, trajectory-sensing problem.  If one has a set of detectors, a trajectory is a subset of those detectors.  Different subsets can fire, and one wants to discriminate the resulting output states corresponding to different subsets, that is, trajectories, without error.  The conditions on trajectories, input sensor states, and interaction strengths were analyzed in \cite{chuang1,chuang2}.
	
Here we will analyze how well one can discriminate trajectories, which we shall call patterns, in the minimum-error regime in one and two dimensional detector arrays.  We will be considering the small interaction strength regime, which means that the output states we are discriminating will not be orthogonal.  Consequently, there will be a probability of error, which we will minimize.  This is an extension of our work in \cite{hillery,zhan,ali}. 

We will begin with a discussion of our model for a detector array and the mathematical formalism for detecting patterns that are shifted versions of each other.  We will proceed by looking at the simple case of a one-dimensional array consisting of four detectors and using it to examine the effect of different initial states of the array on the probability of successfully detecting which pattern of excited detectors is present.  This analysis of initial states is continued in the next section in order to compare two types of initial states for larger arrays of detectors.  We then present general results for one of these initial states for a one dimensional array of $N$ detectors where we wish to detect a contiguous block of excited detectors.  Finally, we generalize our results to the case of two-dimensional detector arrays.

\section{Basic model and shift operator}

Suppose we have a line of detectors numbered $1$ through $N$.  Each detector is a qubit.  The interaction between a detector and the environment is described by a unitary operator, $U$, whose eigenvectors we will take to be the computational basis elements $\{ |0\rangle , |1\rangle\}$.  In particular, $U|0\rangle = e^{i\theta}|0\rangle$ and $U|1\rangle = e^{-i\theta}|1\rangle$.  If a detector has fired, that means the operator $U$ has been applied to its state, if it has not fired, its state is unchanged.  We would like to be able to distinguish between shifted firing patterns of detectors.  What this means is the following.  Suppose some subset of the detectors has fired.  This creates a pattern of fired and not-fired detectors.  Now shift this pattern one step to the right.  Keep doing this, treating the line of detectors as if they are on a ring, and we will create $N$ patterns each corresponding to a different state.  We want to distinguish these states.  To be more explicit, suppose we have $3$ detectors initially in the state $|\psi_{in}\rangle$. One possibility is that we want to discriminate among the states $(U\otimes I\otimes I)|\psi_{in}\rangle$, $(I\otimes U\otimes I)|\psi_{in}\rangle$, and $(I\otimes I\otimes U)|\psi_{in}\rangle$.  Another is that we want to discriminate among $(U\otimes U\otimes I)|\psi_{in}\rangle$, $(I\otimes U\otimes U)|\psi_{in}\rangle$, and $(U\otimes I\otimes U)|\psi_{in}\rangle$.

Something that will be useful in our quest is the shift operator on $N$ qubits.  It can be defined by its action on the basis states $|j_{1}\rangle_{1} |j_{2}\rangle_{2} |j_{3}\rangle_{3} \ldots |j_{N}\rangle_{N}$, where $j_{k} = 0,1$.  The shift operator, $S$, acts on the basis states as
\begin{eqnarray}
S |j_{1}\rangle_{1} |j_{2}\rangle_{2} |j_{3}\rangle_{3} \ldots |j_{N}\rangle_{N} & = & |j_{N}\rangle_{1} |j_{1}\rangle_{2} |j_{2}\rangle_{3} |j_{3}\rangle_{4} \ldots \nonumber \\
&& |j_{N-1}\rangle_{N} .
\end{eqnarray}
This operator can be used to shift the pattern of firing detectors.  For example, for three qubits we have
\begin{equation}
S(U\otimes I \otimes I)S^{-1} = I \otimes U \otimes I .
\end{equation}
This can be seen by noting
\begin{eqnarray}
S(U\otimes I \otimes I)S^{-1} |\phi_{1},\phi_{2},\phi_{3}\rangle & = & S(U\otimes I \otimes I)  |\phi_{2},\phi_{3},\phi_{1}\rangle \nonumber \\
& = & S |U\phi_{2},\phi_{3},\phi_{1}\rangle \nonumber \\
& = & |\phi_{1},U\phi_{2},\phi_{3}\rangle .
\end{eqnarray}
Let the initial state of the sensors be $|\psi_{in}\rangle$ and the basic pattern of firing detectors be described by $V$, which is a tensor product of $U$ and $I$ operators.  Then the first of our states to be distinguished is $|\eta_{0}\rangle = V|\psi_{in}\rangle$, and the remainder are of the form $|\eta_{n}\rangle = S^{n}VS^{-n}|\psi_{in}\rangle$ for $n=1,2, \ldots N-1$.  Now we are going to make an assumption, we will assume that $|\psi_{in}\rangle$ is a symmetric state, that is $S|\psi_{in}\rangle = |\psi_{in}\rangle$.  That means that $|\eta_{n}\rangle = S^{n}V|\psi_{in}\rangle$.  We can now apply a general result described below.

In order to find which detector fired, we need to be able to discriminate among the states $\{|\eta_{j}\rangle \}$.  In general, these states will not be orthogonal and cannot be discriminated perfectly.  We will make use of the minimum-error strategy.  For $N$ states, we have an $N$-element POVM with elements $\Pi_{j}$.  The $\Pi_{j}$ are positive and satisfy $\sum_{j=1}^{N}\Pi_{j}=I$.  The probability that if we are given the state $|\eta_{k}\rangle$ that detector $j$ clicks is $\langle \eta_{k}|\Pi_{j}|\eta_{k}\rangle$.  Detector $j$ clicking is supposed to correspond to the detection of the state $|\eta_{j}\rangle$, so our probability of successfully identifying a given state, if each of the states is equally probable, is
\begin{equation}
P_{s}=\frac{1}{N}\sum_{j=0}^{N-1}\langle \eta_{j}|\Pi_{j}|\eta_{j}\rangle .
\end{equation}
In minimum-error state discrimination, we seek to find a POVM that maximizes this success probability, and, consequently, minimizes the probability of making a mistake.  The solution to the problem is known in complete generality for two states, but only in special cases for more than two states.  One of these special cases is that of symmetric states, in which the states are related by a single unitary operator, and we shall make use of this case \cite{ban,eldar}.

We have a set of equally probable states $\{ |\eta_{j}\rangle = S^{j}|\eta_{0}\rangle \, | \, j=0,1,\ldots N-1\}$, where $S$ is a unitary operator with the property that $S^{N}=I$.  Because $S^N= I$, the eigenvalues of $S$ are the $N$-roots of unity, and $S$ can be expressed in terms of its eigenstates, $\{ |u_{j}\rangle \}$
\begin{equation}
S=\sum_{j=0}^{N-1}{\rm e}^{2\pi ij/ N}|u_{j}\rangle\langle u_{j}| .
\end{equation}
The state $|\eta_0\rangle$ can be expressed as
\begin{equation}
|\eta_0\rangle=\sum_{j=0}^{N-1} d_j|u_j\rangle .
\end{equation}
The optimal success probability of distinguishing the states is given by (see the Appendix for a discussion)
\begin{equation}
P_{s} = \frac{1}{N}\left(\sum_{j=0}^{N-1}|d_j |\right)^2 .
\end{equation} 

\section{Four detectors}
Let's apply this to $4$ detectors (qubits) with $V=U\otimes I^{\otimes 3}$.  We first need to find the eigenstates of $S$.  Since $S^{4}=I$, its eigenvalues are $e^{2n\pi i/4} = e^{in\pi /2}$ for $n=0,1,2,3$.  The eigenstates fall into groups labeled by their Hamming weights when expressed in the computational basis.  The states $|0000\rangle$ and $|1111\rangle$ (Hamming weights $0$ and $4$, respectively) both correspond to eigenvalue $1$.  The eigenstates in the Hamming weight $1$ and $3$ sectors are
\begin{eqnarray}
|v_{1n}\rangle & = & \frac{1}{2} (|0001\rangle + e^{in\pi /2} |0010\rangle + e^{in\pi} |0100\rangle \nonumber \\
&& + e^{3ni\pi /2} |1000\rangle ) \nonumber \\
|v_{3n}\rangle & = & \frac{1}{2} (|1110\rangle + e^{in\pi /2} |1101\rangle + e^{in\pi} |1011\rangle \nonumber \\ 
&& + e^{3ni\pi /2} |0111\rangle) .
\end{eqnarray}
for $n=0,1,2,3$, where $S|v_{1n}\rangle = e^{in\pi /2} |v_{1n}\rangle$, and similarly for $|v_{3n}\rangle$.  The Hamming weight $2$ sector consists of two groups
\begin{eqnarray}
|v_{2n}\rangle & = & \frac{1}{2} (|0011\rangle + e^{in\pi /2} |0110\rangle + e^{in\pi} |1100\rangle \nonumber \\
&& + e^{3ni\pi /2} |1001\rangle  ,
\end{eqnarray}
for $n=0,1,2,3$ and 
\begin{equation}
|v_{2\pm}\rangle = \frac{1}{\sqrt{2}}(|0101\rangle \pm |1010\rangle ) ,
\end{equation}
which correspond to eigenvalues $\pm1$.

As an illustration, let us look at a choices for $|\psi_{in}\rangle$ in the Hamming weight two sector for a single excitation.  We require that $S|\psi_{in}\rangle = |\psi_{in}\rangle$.  In this sector, the most general state satisfying this condition is $|\psi_{in}\rangle = \alpha |v_{20}\rangle + \beta |v_{2+}\rangle$.  We now have that
\begin{eqnarray}
|\eta_{0}\rangle & = & V|\psi_{in}\rangle \nonumber \\
& = & \alpha \sum_{j=0}^{3} |v_{2j}\rangle\langle v_{2j}|V|v_{20}\rangle + \beta \sum_{j=\pm}  |v_{2j}\rangle\langle v_{2j}|V|v_{2+}\rangle . \nonumber \\
\end{eqnarray}
We now need to expand $|\eta_{0}\rangle$ in terms of the eigenstates of $S$ in order to find the coefficients $d_{j}$.  The relevant eigenstates appearing in the above sum, up to normalization, are
\begin{eqnarray}
\label{4eigen}
\lambda = 1 & \hspace{5mm}& \alpha |v_{20}\rangle\langle v_{20}|V|v_{20}\rangle + \beta |v_{2+}\rangle \langle v_{2+}|V|v_{2+}\rangle \nonumber \\ 
\lambda = i & \hspace{5mm} &  \alpha |v_{21}\rangle\langle v_{21}|V|v_{20}\rangle \nonumber \\
\lambda = -1 & \hspace{5mm} & \alpha |v_{22}\rangle\langle v_{22}|V|v_{20}\rangle + \beta |v_{2-}\rangle \langle v_{2-}|V|v_{2+}\rangle \nonumber \\
\lambda = -i & \hspace{5mm} &  \alpha |v_{23}\rangle\langle v_{23}|V|v_{20}\rangle .
\end{eqnarray}
We have already calculated the matrix elements $\langle v_{2j}|V|v_{20}\rangle$, and we now find that
\begin{equation}
\langle v_{2+}|V|v_{2+}\rangle = \cos\theta \hspace{5mm} \langle v_{2-}|V|v_{2+}\rangle = i\sin\theta .
\end{equation}
This gives us
\begin{eqnarray}
|d_{0}| = |\cos\theta | & \hspace{5mm} & |d_{2}| = |\beta \sin\theta | \nonumber \\
|d_{1}| = \frac{|\alpha |}{\sqrt{2}} |\sin\theta | & \hspace{5mm} & |d_{3}| = \frac{|\alpha |}{\sqrt{2}} |\sin\theta | ,
\end{eqnarray}
and
\begin{equation}
\frac{1}{4}\left( \sum_{j=0}^{3} |d_{j}|\right)^{2} = \frac{1}{4} [ |\cos\theta | + (|\beta |+ \sqrt{2} |\alpha |) |\sin\theta | ]^{2} .
\end{equation}
We want to choose $|\alpha |$ and $|\beta |$ to maximize this expression, subject to $|\alpha |^{2} + |\beta |^{2} = 1$.  Setting $\alpha = \cos\phi$ and $\beta = \sin\phi$, for $0\leq \phi \leq \pi /2$, we find that $\sin\phi + \sqrt{2}\cos\phi$ achieves a maximum value of $\sqrt{3}$ when $\tan\phi = 1/\sqrt{2}$.  With this choice, 
\begin{equation}
\frac{1}{4}\left( \sum_{j=0}^{3} |d_{j}|\right)^{2} = \frac{1}{4} [ |\cos\theta | + \sqrt{3} |\sin\theta | ]^{2} .
\end{equation}
Note that if $\tan\phi = 1/\sqrt{2}$, then $\sin\phi = \sqrt{1/3}$ and $\cos\phi = \sqrt{2/3}$ so that $|\psi_{in}\rangle$ is just the Dicke state of 4 qubits of Hamming weight 2.  This is consistent with our previous results where we found that the best initial state for detecting a single fired detector in a set of $N$ detectors was a Dicke state of $N$ qubits ($N$ is even) with a Hamming weight of $N/2$ \cite{hillery,zhan}.

Now let's look at the case of two excitations.  In that case $V=U^{\otimes2}\otimes I^{\otimes 2}$.  The relevant eigenstates are given by Eq.\ (\ref{4eigen}) with the two-excitation operator $V$.  We find 
\begin{eqnarray}
\lambda = 1 & \hspace{5mm}& \frac{\alpha}{2}(1+\cos 2\theta ) |v_{20}\rangle + \beta |v_{+}\rangle \nonumber \\
\lambda = i & \hspace{5mm} & \frac{i\alpha}{2} \sin 2\theta |v_{21}\rangle \nonumber \\
\lambda = -1 & \hspace{5mm} & \frac{\alpha}{2} (\cos 2\theta -1) |v_{22}\rangle \nonumber \\
\lambda = -i & \hspace{5mm} & \frac{i\alpha}{2} \sin 2\theta |v_{23}\rangle .
\end{eqnarray}
This gives 
\begin{eqnarray}
|d_{0}| = \left[ \frac{|\alpha |^{2}}{4}(1+\cos 2\theta )^{2} + |\beta |^{2}\right]^{1/2} & \hspace{2mm} & |d_{1}|= \frac{|\alpha |}{2}|\sin 2\theta | \nonumber \\
|d_{2}| = \frac{|\alpha |}{2} (1-\cos 2\theta ) & \hspace{2mm} & |d_{3}|  = \frac{|\alpha |}{2} |\sin 2\theta | .\nonumber \\ &&
\end{eqnarray}
This results in a success probability of 
\begin{eqnarray}
\label{4-qubit-ent}
P_{s} & = & \frac{1}{4} \left\{ \left[ \frac{|\alpha |^{2}}{4}(1+\cos 2\theta )^{2} + |\beta |^{2}\right]^{1/2} + |\alpha \sin 2\theta | \right. \nonumber \\
&&\left. + \frac{|\alpha |}{2} (1-\cos 2\theta ) \right\}^{2} .
\end{eqnarray}
We can use this expression to find the success probabilities for the Dicke state, $\alpha = \sqrt{2/3}$, or what we shall call the contiguous state, $\alpha = 1$, that is the superposition of Hamming weight two states with two adjacent ones.  The results are plotted in Fig.\ 1.

\begin{figure}[h]
\centering
\includegraphics[width=\columnwidth]{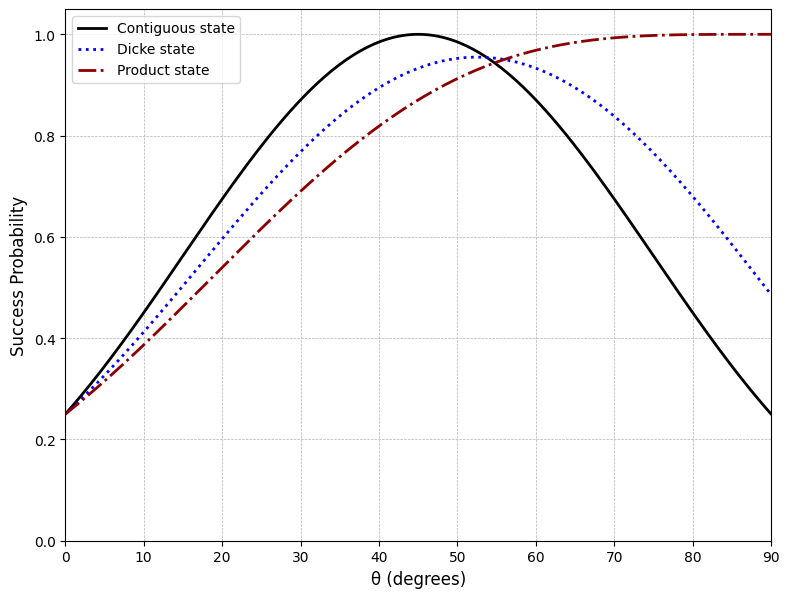}
\caption{Success probabilities of discrimination on a four-qubit sensor array with two excitations, for various initial states.}
\end{figure}

Finally, let's see what happens if $|\psi_{in}\rangle$ is a product state.  The single-qubit state, $|\psi\rangle$ for which the magnitude of the overlap between $|\psi\rangle$ and $U|\psi\rangle$ is minimized is $|+x\rangle$ ($|-x\rangle$ works as well) where $|\pm x\rangle = (|0\rangle \pm |1\rangle )/\sqrt{2}$.  This choice guarantees that $|\psi\rangle$ and $U|\psi\rangle$ are maximally distiguishable.  We will choose $|\psi_{in}\rangle = |+x\rangle ^{\otimes 4}$ and look at the two excitation case, so that $V=U^{\otimes 2}\otimes I^{\otimes 2}$.  Let $Q_{n}$ be the projector onto the eigenvalue $e^{in\pi /2}$ subspace in the four-qubit Hilbert space.  We then have that $|d_{n}|^{2}=\langle\psi_{in} |V^{-1}Q_{n}V|\psi_{in}\rangle$, or, more explicitly

\vspace{-.4cm}

\begin{eqnarray}
|d_{0}| & = & \frac{1}{2} (1+ \cos^2 \theta ), \quad |d_{2}| = \frac{1}{2} \sin ^2 \theta ,\nonumber \\
|d_{1}| & = & |d_{3}| = \frac{1}{2} \sin \theta \sqrt{1+ \cos^2 \theta}
\end{eqnarray}

Putting these expressions into the success probability, we find that for small $\theta$ the success probability for the product state is $(1/4)(1+\sqrt{2}\theta )^{2}$ and that for the contiguous state, Eq.\ (\ref{4-qubit-ent}) with $\alpha = 1$ and $\beta = 0$, is $(1/4)(1+2\theta )^{2}$, which is better.  All three results, Dicke, contiguous, and product, are plotted in Fig. 1.  We see that for small values of $\theta$, corresponding to a weak detector-signal interaction, the contiguous state performs best.  Both it and the Dicke state are entangled, while the product state is not.  This shows that for four detectors with two adjacent detectors firing, entanglement helps.

\section{N qubits}
\subsection{Two excited detectors} 
In some cases we can extend our results to $N$-qubit states.  We will start by looking at the case in which there are two adjacent excited detectors, and we will choose for the initial state of the detectors the Hamming weight $2$ state
\begin{equation}
|e_{0}\rangle = \frac{1}{\sqrt{N}} \sum_{k=0}^{N-1} S^{k}|110\ldots 0\rangle ,
\end{equation}
which is a superposition of shifted states.  The relevant eigenstates of $S$ for this case are given by
\begin{equation}
|e_{j}\rangle = \frac{1}{\sqrt{N}} \sum_{k=0}^{N-1} e^{2\pi ijk/N} S^{k}|110\ldots 0\rangle ,
\end{equation}
where $j=0,1,\ldots N-1$ and $S|e_{j}\rangle = e^{-2\pi ij/N} |e_{j}\rangle$.  We now have $V=U\otimes U\otimes I^{\otimes (N-2)}$, and 
\begin{eqnarray}
d_{j} & = & \frac{1}{N} \sum_{k=0}^{N-1} e^{-2\pi i jk/N} \langle 110\ldots 0|S^{-k}VS^{k}|110\ldots 0\rangle \nonumber \\
& = & \frac{1}{N}\left[ e^{-2i\theta} + e^{-2\pi ij/N} + e^{2i\theta} \sum_{k=2}^{N-2}e^{-2\pi ijk/N} \right. \nonumber \\
&&\left. + e^{-2\pi i(N-1)/N} \right] .
\end{eqnarray}
This can be simplified to

\begin{equation}
d_j = \delta_{j0} \ e^{2i\theta}  - \frac{4i \sin \theta}{N} \left( \cos \theta + e^{i\theta} \cos 2 \phi_j  \right)   
\end{equation}
where $\phi_j = \frac{ \pi j}{N}$.
\vspace{.2cm}

We found that in the case of a single detector firing out of a total of $N$ detectors, the best initial state for identifying that detector was the Dicke state, $|N;N/2\rangle$, i.e.\ the equal superposition of all basis states with $N/2$ ones.  We have seen above that for four qubit detectors and two detectors firing, the Dicke state $|2;2\rangle$ is not optimal.  It would be useful to know whether the result for four detectors holds for a larger number of detectors.  This suggests looking at $N$ detectors with two adjacent ones firing, and comparing the behavior of the Dicke state to that of the state that is a superposition of shifted states.  The idea here is to look at the set of states $\{ |v_{n}\rangle = S^{n}(U\otimes U \otimes I^{\otimes (N-2)}|N;l\rangle\, |\, n=0, 1, \ldots N-1\}$, where $|N;l\rangle$ is the Dicke state with $l$ ones.  One way of finding the optimal minimum-error POVM for this case is to make use of the pretty good measurement that discriminates these states \cite{hausladen,eldar2}.

We are going to need the overlaps between the states in this set.  There are three cases. The first is the trivial one, the overlap of a state with itself is one.  The next one is when there is no overlap between the locations of the $U$ operators in the two states.  In this case, the overlap of the states will be given by 
\begin{equation}
a=\langle N;l|(U\otimes U\otimes U^{-1}\otimes U^{-1} \otimes I^{\otimes (N-4)} |N;l\rangle .
\end{equation}
Finding this requires counting the number of terms that give us $e^{\pm 4i\theta}$, $e^{\pm 2i\theta}$, and $1$.  The result is
\begin{eqnarray}
a & = & \frac{(N-4)!}{N!} \{ 2l(l-1)(N-l)(N-l-1) \cos4\theta \nonumber \\
&&+ 4l(N-l)[ (N-l-1)(N-l-2) + (l-1)(l-2)] \nonumber \\
&& \cdot \cos 2\theta +  4l(l-1)(N-l)(N-l-1) \nonumber \\
&&+ (N-l)(N-l-1)(N-l-2)(N-l-3) \nonumber \\
&&+ l(l-1)(l-2)(l-3) \} 
\end{eqnarray}
The third case is when two of the $U$ operators overlap.  In this case the overlap between the states is
\begin{equation}
b = \langle N;l|(U\otimes I\otimes U^{-1}\otimes I^{\otimes (N-3)} |N;l\rangle ,
\end{equation}
and in this case one has to count the terms that give $e^{\pm 2i\theta}$ and $1$.  The result is
\begin{equation}
b = \frac{1}{N(N-1)} [ 2l(N-l) \cos 2\theta + (N-l)(N-l-1) + l(l-1) ] .
\end{equation}

In our case the POVM for the pretty good measurement is given by
\begin{equation}
\Pi_{j} = \frac{1}{N} \rho^{-1/2}|v_{j}\rangle\langle v_{j}|\rho^{-1/2} ,
\end{equation}
for $j=1,2,\ldots N$ and $\rho = (1/N)\sum_{n=1}^{N} |v_{n}\rangle\langle v_{n}|$.  The success probability for this measurement is
\begin{equation}
P_{s}=\frac{1}{N} \sum_{j=1}^{N} \langle v_{j}|\Pi_{j}|v_{j}\rangle = \frac{1}{N^{2}}\sum_{j=1}^{N} |\langle v_{j}|\rho^{-1/2}|v_{j}\rangle |^{2} .
\end{equation}

The problem now is to find $\rho^{-1/2}$ and that requires diagonalizing $\rho$.  To do this we will use the non-orthogonal basis $\{ |v_{n}\rangle \, |\, n=1,2,\ldots N\}$ and write the eigenvectors as $|x\rangle = \sum_{j=1}^{N} x_{j}|v_{j}\rangle$.  The eigenvalue equation becomes
\begin{equation}
\rho |x\rangle = \frac{1}{N}\sum_{n=1}^{N}\sum_{j=1}^{N} |v_{n}\rangle x_{j} \langle v_{n}|v_{j}\rangle = \lambda \sum_{j=1}^{N} x_{j}|v_{j}\rangle .
\end{equation}
This implies that 
\begin{equation}
\frac{1}{N} \left[ b(x_{n-1} + x_{n+1}) + x_{n} + a \sum_{j\neq \{ n-1,n+1\}} x_{j} \right] = \lambda x_{n} ,
\end{equation}
where the additions and subtractions in the subscripts are modulo $N$.  This can be written in matrix form as
\begin{equation}
\frac{1}{N} \left( \begin{array}{cccccccc} 1 & b & a & a & \ldots & a & a & b \\ b & 1 & b & a & \ldots & a & a & a \\ \vdots &&&&&&& \vdots \\ b& a & a & a & \ldots & a & b & 1 \end{array}\right) \left( \begin{array}{c} x_{1} \\ x_{2} \\ \vdots \\ x_{N} \end{array} \right) = \lambda \left( \begin{array}{c} x_{1} \\ x_{2} \\ \vdots \\ x_{N} \end{array} \right) .
\end{equation}
Note that if we call the matrix $M$, its matrix elements are given by $M_{jk}=\langle v_{j}|v_{k}\rangle$.  This matrix is a circulant matrix, and this makes its eigenvalues and eigenvectors simple to find.  The $j^{\rm th}$ eigenvector is
\begin{equation}
|e_{j}\rangle = A_{j} \left( \begin{array}{c} 1 \\ \omega^{j} \\ \omega^{2j} \\ \vdots \\ \omega^{(N-1)j} \end{array} \right) ,
\end{equation}
where $\omega = e^{2\pi i/N}$, and the corresponding eigenvalue is 
\begin{eqnarray}
\lambda_{j} & = & \frac{1}{N}(1+ b(\omega^{j} + \omega^{-j}) + a \sum_{t=p}^{N-2} \omega^{jp} ) \nonumber \\
&=& a\,\delta_{j0}  + \frac{1}{N}\!\left[ (1 - a) + 2(b - a)\cos (2\phi_j) \right] 
\end{eqnarray}
where, as before, we have used $\phi_j = \frac{\pi j}{N}$. In order to normalize the eigenstates, we have that
\begin{eqnarray}
1 & = & \langle e_{j}|e_{j}\rangle = |A_{j}|^{2} \sum_{p=1}^{N}\sum_{p^{\prime}=1}^{N} \omega^{(p^{\prime}-p)j}\langle v_{p}|v_{p^{\prime}}\rangle \nonumber \\
& = & |A_{j}|^{2} \sum_{p=1}^{N} \omega^{-jp} \omega^{jp} N\lambda_{j} = |A_{j}|^{2} N^{2} \lambda_{j} ,
\end{eqnarray}
where we used $\sum_{p^{\prime}=1}^{N} M_{pp^{\prime}}\omega^{jp^{\prime}} = \sum_{p^{\prime}=1}^{N} \langle v_{p}|v_{p^{\prime}}\rangle \omega^{jp^{\prime}} = \omega^{jp}N\lambda_{j}$.  This gives us that $A_{j}=1/(N\sqrt{\lambda_{j}} )$.

We can now find an explicit expression for the success probability.  We have that
\begin{equation}
\langle v_{n}|\rho^{-1/2}|v_{n}\rangle = \sum_{p=0}^{N-1} \lambda_{p}^{-1/2} |\langle v_{n}|e_{p}\rangle |^{2} ,
\end{equation}
and
\begin{equation}
\langle v_{n}|e_{p}\rangle = \frac{1}{N\sqrt{\lambda_{p}}} \sum_{p^{\prime}=1}^{N} \omega^{p^{\prime}p}\langle v_{n}|v_{p^{\prime}}\rangle = \omega^{pn} \sqrt{\lambda_{p}} .
\end{equation}
Therefore,
\begin{equation}
\langle v_{n}|\rho^{-1/2}|v_{n}\rangle = \sum_{p=0}^{N-1} \sqrt{\lambda_{p}} ,
\end{equation}
and finally,
\begin{equation} 
P_{s} =\frac{1}{N^{2}}\sum_{n=1}^{N} |\langle v_{n}|\rho^{-1/2}|v_{n}\rangle |^{2} = \frac{1}{N} \left(\sum_{p=0}^{N-1}\sqrt{\lambda_{p}} \right)^{2} .
\end{equation}


We can now compare the results obtained for the two choices of input state.  In Fig.\ 2 we plot the success probabilities for ten detectors when two adjacent ones fire.  One curve is for the Dicke state of Hamming weight 5.  The other is for the contiguous state (superposition of shifted states) of Hamming weight 2.  We did look at different Hamming weights for the contiguous state, but two was optimal in this case.  We see that the Dicke state outperforms the contiguous state, but for small $\theta$, the difference is not large.  Since the Dicke state is considerably more complicated than the contiguous state, and the performance difference in the low-$\theta$ range is not great, it may prove to be the more reasonable alternative.


\subsection{Generalization for arbitrary number of qubits and excitations}

While the fully general case is complicated to analyze, a closed‑form expression can be found if we restrict ourselves to the particular class of states that we will call 'contiguous initial states' defined by:
\begin{equation}
\left|\Psi_{\mathrm{in}}\right\rangle=\frac{1}{\sqrt{N}} \sum_{k=0}^{N-1} S^k|\underbrace{11 \ldots 1}_l {0000}\rangle
\end{equation}

Having fixed the form of these initial states, the most general scenario would have all arbirtrary parameters $l$,$m$ and $N$, which represent the Hamming weight of the contiguous states, the number of excitations (the length of the patterns) and the number of qubits respectively. The corresponding interaction operator is:
\begin{equation}
    V = U^{\otimes m} \otimes I^{\otimes (N-m)}
    \label{eq:interaction_operator_general}
\end{equation}

The coefficients $d_j$ are projections of the post-interaction state onto the eigenvectors of the shift operator (see Appendix):
\begin{equation}
    d_j = \langle e_j | V| \Psi_{\text{in}} \rangle
    \label{eq:dp_definition_general}
\end{equation}

Since our interaction operator $V$ leaves the Hamming weight invariant, we realize the only non-zero 
contributions in the inner product come from the eigenvectors that belong to the subspace spanned by the 
component basis states of the initial contiguous state. Thus the relevant eigenvectors to our calculations are of the form
\begin{equation}
    |e_j\rangle = \frac{1}{\sqrt{N}} \sum_{n=0}^{N-1} e^{2 \pi i j \frac{n}{N}} S^n | \underbrace{11\ldots1}_{l}000\ldots0 \rangle
    \label{eq:eigenvectors_detailed_general}
\end{equation}

To proceed, we note that the interaction operator \eqref{eq:interaction_operator_general} affects only the phases, the qubits affected by $U$ either gain positive or negative phase $\theta$. The total phase corresponding to a particular basis state depends on the position of the ones and zeroes in the array, along with the parameters $m,l$. For example, take the simple case of $N, m,l = 6,2,3$. The interaction operator becomes $V = U \otimes U \otimes I^{\otimes 4}$.  The phases corresponding to some of the possible basis states are 
\begin{equation}
\begin{aligned}    
& V|111000\rangle = \\ \quad & \qquad  U \otimes U \otimes I\otimes I \otimes I\otimes I |111000\rangle =  e^{-2i\theta}|111000\rangle, \\ &V|011100\rangle = \\ \quad & \qquad U \otimes U \otimes I\otimes I \otimes I\otimes I |011100\rangle = e^{-i\theta}|011100\rangle, \\
&V|001110\rangle = \\ \quad & \qquad U \otimes U \otimes I\otimes I \otimes I\otimes I  |001110\rangle = |001110\rangle.\\
\end{aligned}
\end{equation}
This example illustrates that for fixed $m,l$ parameters, the total phase is determined by the number of ones and zeros that overlap with the positions  of the $U$ operators in the full interaction operator $V$. 

In general, the possible overlaps can be grouped into four categories: full overlaps with (\romannumeral 1)  zeros (\romannumeral 2) ones, and partial overlap with (\romannumeral 3) ones and then zeros (\romannumeral 4) zeros and then ones. In this way we can work out the phase generated by the interaction depending on how much the state is shifted i.e., as a function of the index $k$.

We will restrict ourselves to the case when $m \leq l \leq N$, as the other case can be found analogously.

\vspace{1pt}

After substituting the previously given expressions for $\langle e_j|$, $V$ and $|\Psi_{\text{in}} \rangle$, the double sum in \eqref{eq:dp_definition_general} is reduced to a single sum due to the orthogonality of the computational basis states. 
The resulting series consists of four sub‑series of different phases depending on the type of overlap.
\vspace{-9pt}
\begin{equation}
    d_j = \sum_{k=0}^{N-1} \frac{\omega_j^{-k}}{\sqrt{N}} e^{i\theta f(k)}
\end{equation}
where \(\omega_j = e^{2\pi i j/N} := \omega \) , where for notational simplicity we have dropped the subscript on $\omega$, and

\[
f(k) =
\begin{cases} 
2k - m, &  1 \leq k \leq m - 1, \\
m, &  m \leq k \leq N - l, \\
m + 2N - 2l - 2k, &  N - l + 1 \leq k \leq N - l + m - 1, \\
-m, &  N - l + m \leq k \leq N.
\end{cases}
\]

\begin{equation*}
    \begin{split}
    d_j &= \frac{1}{N}\Bigg( \sum_{k=1}^{m-1} + \sum_{k=m}^{N-l} + \sum_{k=N-l+1}^{N-l+m-1}  + \sum_{k=N-l+m}^{N}  \Bigg)\omega^{-k} e^{i\theta f(k)} \\
    &= \frac{1}{N}(S_1 + S_2 + S_3 + S_4)
    \end{split}
\end{equation*}

\vspace{.15cm}

Define $q = \omega^{-1} e^{2i\theta}$ and $r = \omega^{-1} e^{-2i\theta}$. We evaluate each term below. 

\begin{equation*}
    S_1 = \sum_{k=1}^{m-1} \omega^{-k} e^{i\theta (2k - m)} = e^{-i\theta m} \sum_{k=1}^{m-1} q^k = e^{-i\theta m} \left(\frac{q - q^{m}}{1 - q}\right)
\end{equation*}

\vspace{-.3cm}
\begin{equation*}
\label{S2}
\begin{aligned}
    S_2 = \sum_{k=m}^{N-l} \omega^{-k} & e^{i\theta m} = e^{i\theta m} \sum_{k=m}^{N-l} \omega^{-k}  \\ 
    & =  \omega^{-m} e^{i m \theta} \left(\frac{\omega - \omega^{l+m}}{\omega - 1}\right)  \qquad \qquad \quad
\end{aligned}
\end{equation*}

\vspace{-.3cm}

\begin{equation*}
\label{S3}
\begin{aligned}
S_3 &= \sum_{k=N-l+1}^{N-l+m-1} \omega^{-k} e^{i\theta (2N - 2l - 2k + m)} \\
& = e^{i\theta ( 2N - 2l + m)} \sum_{k=N-l+1}^{N-l+m-1} r^k = \omega^l e^{i m \theta} \left(\frac{r - r^{m}}{1 - r}\right)
\end{aligned}
\end{equation*}

\vspace{-.5cm}

\begin{equation*}
\label{S4}
\begin{aligned}
S_4 = & \sum_{k=N-l+m}^{N} \omega^{-k} e^{-i\theta m} = e^{-i\theta m} \sum_{k=N-l+m}^{N} \omega^{-k}  \\ 
&= e^{-i m \theta} \omega^{l-m} \left(\frac{\omega - \omega^{m-l}}{\omega - 1}\right) 
\end{aligned}
\end{equation*}

Putting it all together,  we have:
\begin{equation}
    \begin{split}
    N d_j &  =   e^{-im\theta} \cdot \frac{q - q^m}{1 - q} + \omega^{-m} e^{im\theta} \cdot \frac{\omega - \omega^{l+m}}{\omega - 1} \\
    & + \omega^l e^{im\theta} \cdot \frac{r - r^m}{1 - r} + e^{-im\theta} \omega^{l-m} \cdot \frac{\omega - \omega^{m-l}}{\omega - 1}
    \end{split}
\end{equation}

\begin{figure}[t]
  \centering
  \includegraphics[width=\columnwidth]{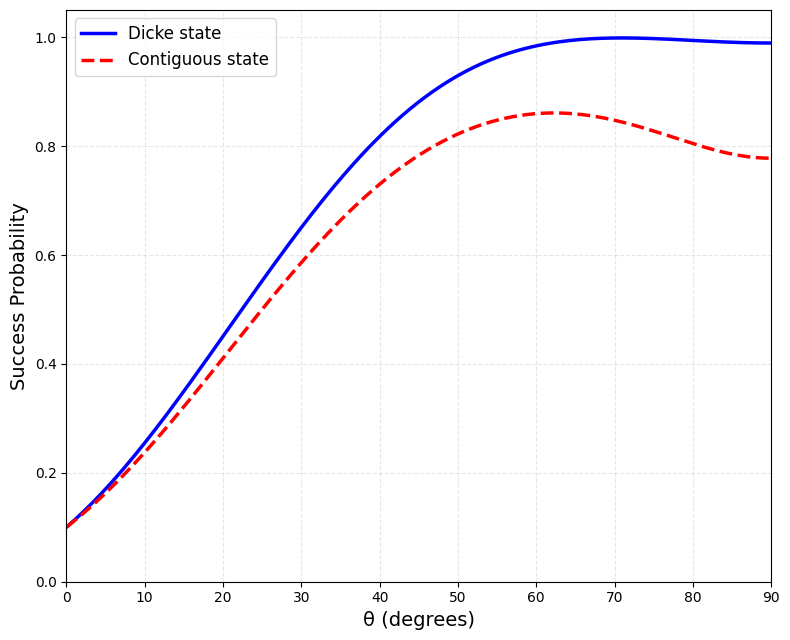}
  \caption{Success probabilities of a 10 qubit sensor array with two excitations initialized in Dicke state of 5 vs contiguous state of 2 Hamming weight.}
  \label{fig:N10 dicke vs ctgs}
\end{figure}

\vspace{.2cm}

Let the half-phase angles of $q$ and $r$ be $\phi_q$ and $\phi_r$. Thus $\phi_q = (\theta - \frac{\pi j}{N})$ and $\phi_r = -\left(\theta + \frac{\pi j}{N}\right)$. Also, define  $\phi_j := \frac{\pi j}{N}$.
We can then rewrite the above expression as
\begin{equation}
\label{fourterm}
\begin{aligned}
     Nd_j ={} & e^{-im\phi_j} \left[ \frac{\sin((m-1)\phi_q)}{\sin\phi_q} + e^{2il\phi_j} \frac{\sin((m-1)\phi_r)}{\sin\phi_r} \right] \\
     & + e^{i(l-m)\phi_j} \left[ (-1)^j e^{im\theta} \frac{\sin((N-l-m+1)\phi_j)}{\sin\phi_j} \right. \\
     & \quad \left. + e^{-im\theta} \frac{\sin((l-m+1)\phi_j)}{\sin\phi_j} \right]
\end{aligned}
\end{equation}

After careful algebraic manipulation \eqref{fourterm} can be equivalently expressed as

\vspace{-.1cm}
\vspace{-.1cm}
\vspace{-.1cm}
\vspace{-.1cm}

\begin{equation}
\label{gneralized ctgs}
\begin{aligned}
 d_{j} =\delta_{j0} e^{im\theta} + \frac{e^{-i m \phi_j}}{N}\left(\frac{\sin \theta}{\sin \phi_j}\right)&\left[\frac{\sin \left(m \phi_q\right)}{\sin \phi_q}-e^{2 i l \phi_j} \frac{\sin \left(m \phi_r\right)}{\sin \phi_r}\right]  
\end{aligned}
\end{equation}

\vspace{.1cm}
\vspace{.1cm}

This is a relatively simple expression for this generalized case of contiguous states. Note that the $d_0$ term is independent of $N$ and of unit magnitude. As the size of our qubit array is increased, the other $d_j$ become smaller being inversely proportional to $N$, which makes $d_0$ increasingly dominant in determining the success probability.


\section{Two-dimensional arrays}

So far, we have only considered detectors in a line, but now we want to examine a square array of detectors.  In order to do so, we are going to need two shift operators, a horizontal shift and a vertical shift.  Our qubit array will be $N\times N$, and we will write our states as tensor products of rows, $|{\rm row\, 0}\rangle \otimes |{\rm row\, 1}\rangle  \otimes \ldots \otimes |{\rm row}\, N-1\rangle$, where each row is a tensor product of $N$ qubits.  Our shift operators are $S_{h}$, which shifts each row one step to the right, and $S_{v}$, which moves each row up one step.  For our initial state we will choose
\begin{equation}
|\psi_{in}\rangle = \frac{1}{N} \sum_{j,k=0}^{N-1} S_{h}^{j} S_{v}^{k} |\phi_{0}\rangle ,
\end{equation}
where $|\psi_{0}\rangle$ is a state in the computational basis.  The operators $S_{h}$ and $S_{v}$ commute, so they have common eigenstates.  The relevant eigenstates are
\begin{equation}
|e_{lm}\rangle = \frac{1}{N} \sum_{j,k=0}^{N-1} e^{-2\pi i jl/N} e^{-2\pi i km/N} S_{h}^{j} S_{v}^{k} |\phi_{0}\rangle  .
\end{equation} 
As before, the operator $V$, which creates the pattern of triggered detectors, is a product of $U$ and identity operators.  The coefficients $d_{lm}$ are given by
\begin{eqnarray}
d_{lm} & = & \langle e_{lm}|V|\psi_{in}\rangle. \nonumber \\
& = & \frac{1}{N^{2}} \sum_{j^{\prime},k^{\prime} =0}^{N-1} \sum_{j,k=0}^{N-1} e^{2\pi i j^{\prime}l/N} e^{2\pi i k^{\prime}m/N} \nonumber \\
&& \langle\phi_{0}| S_{h}^{-j^{\prime}} S_{v}^{-k^{\prime}}VS_{h}^{j} S_{v}^{k}|\phi_{0}\rangle \nonumber \\
&=& \frac{1}{N^{2}} \sum_{j,k=0}^{N-1} e^{2\pi i jl/N} e^{2\pi i km/N} \langle S_{h}^{j} S_{v}^{k} \phi_{0} | V | S_{h}^{j} S_{v}^{k} \phi_{0}\rangle .
\end{eqnarray}
The last line holds if the states $S_{h}^{j}S_{v}^{k}|\phi_{0}\rangle$ are orthogonal to each other (these states are eigenstates of $V$), and in the examples we examine here that will be the case.  As shown in the appendix, the success probability for distinguishing the $N^{2}$ detector states $S_{h}^{j}S_{v}^{k}V|\psi_{in}\rangle$, $j,k=0,1,\ldots N-1$ is
\begin{equation}
P_{s}=\frac{1}{N^{2}} \left( \sum_{j,k=0}^{N-1} | d_{jk}| \right)^{2} .
\end{equation}

Now suppose a $2\times 2$ block of detectors will be excited and we want to find which one.  Our basic unit of excited detectors will be the $2\times 2$ square in the lower left-hand corner of our array, and states with other blocks of shifted detectors will just be shifted versions of this one.  In that case we should choose $V= U^{\otimes 2} \otimes I^{\otimes (N-2)}\otimes U^{\otimes 2} \otimes I^{\otimes (N-2)}\otimes I^{\otimes N(N-2)}$.  We will choose $|\phi_{0}\rangle$ to be a state in the computational basis with ones in our basic unit and zeroes elsewhere, that is $|\phi_{0}\rangle = |1100\ldots 0\rangle \otimes |1100\ldots 0\rangle \otimes |0000\ldots 0\rangle \ldots ||0000\ldots 0\rangle$.  We can then compute the coefficients $d_{lm}$, and we find for $m=l=0$
\begin{equation}
d_{00}  =  \frac{1}{N^{2}}\left[ e^{-4i\theta} + 4(e^{2i\theta} + 1) + (N^{2}-9)e^{4i\theta} + 4\right] .
\end{equation}
For $d_{0m}$ we have
\begin{eqnarray} 
d_{0m} & = & \frac{1}{N^{2}} \left\{ e^{-4i\theta} + 2 + ( 2 + 4e^{2i\theta}) \cos\left(\frac{2\pi m}{N}\right) \right. \nonumber \\
&& \left. - 3 e^{4i\theta}\left[ 1 + 2 \cos\left(\frac{2\pi m}{N}\right) \right] \right\} .
\end{eqnarray}
The expression for $d_{l0}$ is the same as that above but with $m$ replaced by $l$.  Finally, for $l\neq 0$ and $m\neq 0$ we have
\begin{eqnarray}
d_{lm} & = & \frac{1}{N^{2}} \left\{ e^{-4i\theta} + 2 \cos\left(\frac{2\pi l}{N}\right) + 2 \cos\left(\frac{2\pi m}{N}\right) \right. \nonumber \\
&& 2 e^{2i\theta} \left[ \cos\left(\frac{2\pi (l+m)}{N}\right) + \cos\left(\frac{2\pi (l-m)}{N}\right) \right] \nonumber \\
&& \left.-e^{4i\theta} \left[ 1 + 2\cos\left(\frac{2\pi l}{N}\right) \right] \left[ 1 + 2\cos\left(\frac{2\pi m}{N}\right) \right] \right\} . \nonumber \\
\end{eqnarray}

\begin{figure}[htbp]
\centering
\includegraphics[width=0.48\textwidth]{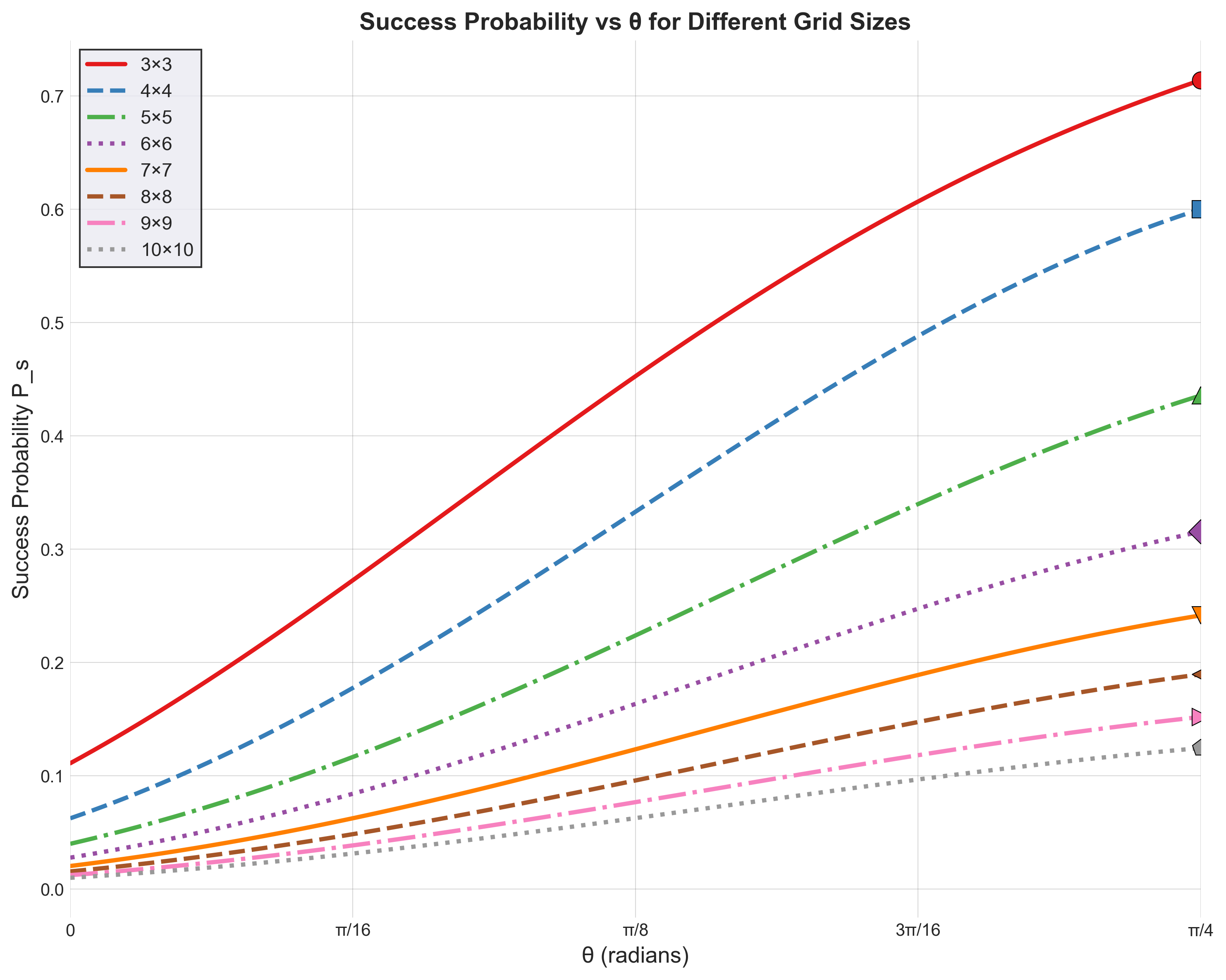}
\caption{Success probability $P_s$ versus interaction strength $\theta$ for $n\times n$ detector grids with $2\times 2$ adjacent block detection.}
\label{fig:ps_vs_theta}
\end{figure}

To vizualize the behavior of the success probability, in Figure~\ref{fig:ps_vs_theta} we plot the success probability $P_s$ as a function of the interaction strength parameter $\theta$ for grid sizes ranging from $3\times 3$ to $10\times 10$. For all grid sizes analyzed, the success probability exhibits a monotonic increase over the range $\theta \in [0, \pi/4]$.  The main observation to emerge from this analysis is that the absolute success probability decreases with increasing grid size, which is expected since the number of possible firing positions scales as $N^2$. For a $3\times 3$ grid there are 9 possible positions, and the maximum of $P_s$ is approximately $0.714$, whereas for a $10\times 10$ grid with 100 positions, the maximum of $P_s$ is $0.125$.

\begin{figure}[htbp]
\centering
\includegraphics[width=0.48\textwidth]{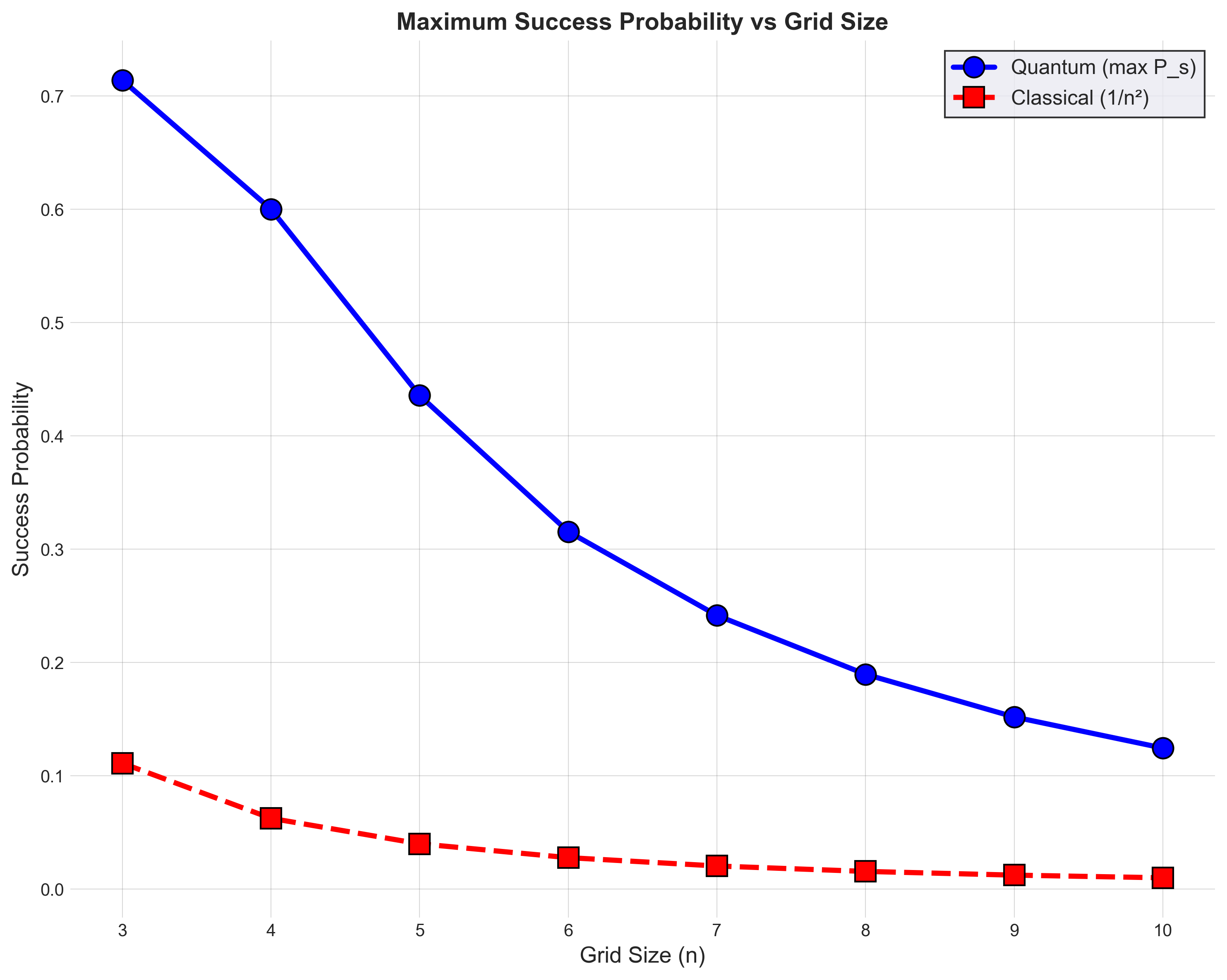}
\caption{Comparison of maximum quantum success probability (blue circles) versus classical random guessing (red squares) as a function of grid size $n$.}
\label{fig:quantum_vs_classical}
\end{figure}

It is useful to compare the maximum quantum success probability to random guessing in order to see the dependence on grid size.  In Figure~\ref{fig:quantum_vs_classical} the maximum quantum success probability (at $\theta = \pi/4$) and the random guessing sucess probability, which is $P_{\text{guess}} = 1/N^2$. The quantum strategy outperforms the random-guessing baseline by a factor that grows with grid size. The widening gap between the curves demonstrates that the relative advantage increases even as absolute probabilities decrease, arising from the collective properties of the entangled initial state $|\psi_{\text{in}}\rangle$ which encodes correlations among all $N^2$ detector positions simultaneously.

Table~\ref{tab:advantage} quantifies this performance improvement by means of the ratio $P_s/P_{\text{guess}}$. The data shows that the advantage grows from $6.43\times$ for $3\times 3$ grids to $12.45\times$ for a $10\times 10$ grid, so that it almost doubles over this range. The growth rate shows signs of saturation for larger grids with diminishing incremental improvements as $N$ increases.

\begin{table}[htbp]
\centering
\caption{Quantum success probability versus guessing probability for $n\times n$ detector grids at $\theta = \pi/4$}
\label{tab:advantage}
\begin{tabular}{ccccc}
\hline\hline
Grid & States & $P_{\text{guess}}$ & $P_s$ & Advantage \\
Size & $(n^2)$ & & & Factor \\
\hline
$3\times 3$ & 9 & 0.111 & 0.714 & $6.43\times$ \\
$4\times 4$ & 16 & 0.063 & 0.600 & $9.60\times$ \\
$5\times 5$ & 25 & 0.040 & 0.436 & $10.89\times$ \\
$6\times 6$ & 36 & 0.028 & 0.316 & $11.36\times$ \\
$7\times 7$ & 49 & 0.020 & 0.242 & $11.84\times$ \\
$8\times 8$ & 64 & 0.016 & 0.190 & $12.14\times$ \\
$9\times 9$ & 81 & 0.012 & 0.152 & $12.31\times$ \\
$10\times 10$ & 100 & 0.010 & 0.125 & $12.45\times$ \\
\hline\hline
\end{tabular}
\end{table}
  
\section{Conclusion}

We have analyzed pattern detection in discrete outcome quantum sensor networks using minimum error state discrimination in both one and two dimensional detector arrays. Our analysis, focusing on the small interaction strength regime where output states are non-orthogonal, reveals several key findings.  First, while Dicke states $|N; N/2\rangle$ are optimal for single detector excitations in one-dimensional arrays, this is not necessarily the case when multiple adjacent detectors are excited. For a small array the contiguous state gave a higher success probability but for a larger array the situation was reversed. For the four detector array, we also showed that both entangled states, the Dicke and the contiguous state, had a higher success probability than the product state. Second, for two-dimensional $N \times N$ arrays, we studied the problem of determining which $2\times 2$ block of detectors fired.  We then compared the quantum success probability to  the probability of guessing the correct answer.  We found that the quantum probability significantly exceeds the guessing probability, with a relative advantage that scales with network size reaching $12.45\times$ for $10 \times 10$ arrays even as absolute success probabilities decrease with the number of possible patterns, $N^2$. Third, the shift operator formalism for symmetric states provides a powerful and tractable tool for calculating success probabilities for arrays in one and two dimensions and for different patterns of firing detectors.

Future work should investigate different excitation patterns, including some that are not contiguous. The same techniques developed here can be applied to three-dimensional detector arrays. Further directions would be to allow the parameter describing the interaction strength, $\theta$, to have a small range of values rather than be fixed, which would result in the states we are trying to discriminate being mixed. Finally, we could begin exploring experimental implementations of these protocols. 

\section*{Appendix}
We start with a set of equally probable states $\{ |\eta_{j}\rangle = S^{j}|\eta_{0}\rangle \, | \, j=0,1,\ldots N-1\}$, where $S$ is a unitary operator with the property that $S^{N}=I$.  As noted in the text, $S$ can be expressed in terms of its eigenstates $\{ |u_{j}\rangle \, |\, j=0,1,\ldots N-1\}$ where $S|u_{j}\rangle e^{2\pi ij/N}|u_{j}\rangle$.  Also, as noted, $|\eta_{0}\rangle$ can be expressed as
\begin{equation}
|\eta_0\rangle=\sum_{j=0}^{N-1} d_j|u_j\rangle ,
\end{equation}
so that 
\begin{equation}
|\eta_k\rangle =S^{k}|\eta_1\rangle = \sum_{j=0}^{N-1} d_j {\rm e}^{2\pi ijk/N} |u_{j}\rangle ,
\end{equation}
which implies, for $0\le l \le N-1$,
\begin{equation}
\langle\eta_{k+l}|\eta_k\rangle = \sum_{j=0}^{N-1} |d_j|^2 {\rm e}^{2\pi ijl/N} ,
\end{equation}
a result independent of $k$. 

We now want to find the minimum-error POVM \cite{bagan}.  Because of the symmetry, we can always choose a POVM to discriminate the states of the form
\begin{equation}
\Pi_j=S^j \Pi (S^\dagger)^j .
\end{equation}
Then, assuming that the states are equally likely, the success probability becomes
\begin{equation}
P_{s}= \langle\eta_0|\Pi|\eta_0\rangle .
\end{equation}
We also have that
\begin{eqnarray}
\label{probbound}
P_{s}&=&\sum_{j,k=0}^{N-1}\Pi_{jk}d^*_j d_k=\left|\sum_{j,k=0}^{N-1}\Pi_{jk}d^*_j d_k\right|\le\sum_{j,k=0}^{N-1}|\Pi_{jk}||d_j| |d_k| \nonumber \\
&\le&\sum_{j,k=0}^{N-1}\sqrt{\Pi_{jj}\Pi_{kk}}|d_j| |d_k|
={1\over N}\sum_{j,k=0}^{N-1}|d_j| |d_k| \nonumber \\
& \leq & {1\over N}\left(\sum_{j=0}^{N-1}|d_j|\right)^2,
\end{eqnarray}
where we have used that positivity implies $|\Pi_{jk}|^2\le|\Pi_{jj}||\Pi_{kk}|$.  In addition, the fact that the POVM elements sum up to the identity implies
\begin{eqnarray*}
\sum_{k=1}^{N}\langle u_j |\Pi_{k}|u_j\rangle & = & \sum_{k=1}^{N} \langle u_j |S^{k-1}\Pi (S^{\dagger})^{k-1}|u_j \rangle \nonumber \\
& = & N\Pi_{jj}=1 .
\end{eqnarray*}
The bound in Eq.\ (\ref{probbound}) is attained by choosing
\begin{equation}
\Pi = \frac{1}{N}\sum_{j,k=0}^{N-1} e^{i(\xi_{j}-\xi_{k})}|u_j \rangle\langle u_k | ,
\end{equation}
where $d_{j}=e^{i\xi_{j}}|d_{j}|$.  The POVM elements, therefore, are 
\begin{equation}
\Pi_{l}=\frac{1}{N}\sum_{j,k=0}^{N-1}e^{2\pi i (j-k)(l-1)/N} e^{i(\xi_{j}-\xi_{k})} |u_j\rangle\langle u_k | ,
\end{equation}
and we note that
\begin{eqnarray}
\sum_{l=1}^{N}\Pi_{l} & = & \frac{1}{N}\sum_{j,k=0}^{N-1}\left[ \sum_{l=1}^{N} e^{2\pi i (j-k)(l-1)/N} \right] \nonumber \\
& & e^{i(\xi_{j}-\xi_{k})}|u_j \rangle\langle u_k | \nonumber \\
& = & \sum_{j=0}^{N-1}|u_j \rangle\langle u_j |=I .
\end{eqnarray} 

The results above allow us to find the optimal POVM for a one-dimensional array of detectors for a given choice of initial state, but if we want to consider a two-dimensional array we need a generalization.  The set of of states we want to discriminate are now given by $\{ |\eta_{jk}\rangle \, | \, j=0, 1, \ldots N-1; k=0,1, \ldots M-1\}$, where $|\eta_{jk}\rangle = U_{1}^{j}U_{2}^{k}|\eta_{00}\rangle$, $U_{1}^{N}=U_{2}^{M}=I$, and $[U_{1},U_{2}]=0$.  The case of discriminating states generated by two unitary operators was first considered by Barnett \cite{barnett}.  The simultaneous eigenstates of $U_{1}$ and $U_{2}$ are $|e_{jk}\rangle$, where
\begin{equation}
U_{1}|e_{jk}\rangle = e^{2\pi ij/N}|e_{jk}\rangle \hspace{5mm} U_{2}|e_{jk}\rangle = e^{2\pi ik/M}|e_{jk}\rangle .
\end{equation} 

We assume our POVM can be expressed as $\Pi_{jk}=U_{1}^{j}U_{2}^{k}\Pi_{00}U_{2}^{-k}U_{1}^{-j}$, so that the success probability is given by
\begin{equation}
P_{s}= \frac{1}{NM} \sum_{j=0}^{N-1}\sum_{k=0}^{M-1} \langle \eta_{jk}|\Pi_{jk}|\eta_{jk}\rangle = \langle\eta_{00}|\Pi_{00}|\eta_{00}\rangle .
\end{equation}
Expanding $|\eta_{00}\rangle$ in terms of the eigenstates $|e_{jk}\rangle$, 
\begin{equation}
|\eta_{00}\rangle = \sum_{j=0}^{N-1}\sum_{k=0}^{M-1} d_{jk} |e_{jk}\rangle ,
\end{equation}
we have
\begin{eqnarray}
P_{s} & = & \sum_{j,k} \sum_{j^{\prime},k^{\prime}} d_{jk}^{\ast} (\Pi_{00})_{jk; j^{\prime}k^{\prime}}d_{j^{\prime},k^{\prime}} \nonumber \\
& \leq &  \sum_{j,k} \sum_{j^{\prime},k^{\prime}} [  (\Pi_{00})_{jk; jk} (\Pi_{00})_{j^{\prime}k^{\prime}; j^{\prime}k^{\prime}} ]^{1/2} |d_{jk}| |d_{j^{\prime},k^{\prime}} | \nonumber \\
& \leq & \frac{1}{MN} \left( \sum_{j,k} | d_{jk}| \right)^{2} ,
\end{eqnarray}
where we have used
\begin{equation}
1 =  \sum_{l=0}^{N-1} \sum{m=0}^{M-1} \langle e_{jk}|\Pi_{lm}|e_{jk}\rangle = MN (\Pi_{00})_{jk;jk} .
\end{equation}
As in the case with one unitary operator, this upper bound can be achieved.

\newpage

\section*{Acknowledgments}
This research was supported by the National Science Foundation under the grant Collaborative Research: NeTS: Medium 2504622.


\newpage
\begin{thebibliography}{99}
		
\bibitem{gorshkov} Z.~Eldredge, M.~Foss-Feig, J.~A.~Gross, S.~L.~Rolston, and A.~V.~Gorshkov, Phys.\ Rev.\ A{\bf 97}, 042337 (2018).
\bibitem{qian} K.~Qian, Z.~Eldredge, Wenchao Ge, G.~Pagano, C.~Monroe, J.~V.~Poeto, and A.~V.~Gorshkov, Phys.\ Rev.\ A {\bf 100}, 042304 (2019).
\bibitem{proctor} T.~J.~Proctor, P.~A.~Knott, and J.~A.~Dunningham, Phys.\ Rev.\ Lett.\ {\bf 120}, 080501 (2018).
\bibitem{Rubio} J.~Rubio, P.~A.~Knott, T.~J.~Proctor, and J.~A.~Dunningham, J.\ Phys.\ A {\bf 53}, 344001(2020).
\bibitem{shapiro} Quntao Zhuang, Zheshen Zhang, and J.~H.~Shapiro, Phys.\ Rev.\ A {\bf 97}, 032329 (2018).
\bibitem{ge}  Wenchao Ge, K.~Jacobs, Z.~Eldredge, A.~V.~Gorshkov, and M.~Foss-Feig, Phys.\ Rev.\ Lett.\ {\bf 121},  043604 (2018).
\bibitem{Zhang} Zheshen Zhang, Quntao, Zhuang, Quantum Sci.\ Technol.\ {\bf 6}, 043001 (2021).
\bibitem{preskill} Quntao Zhuang, J.~Preskill, and Liang Jiang, New J.\ Phys.\ {\bf22} 022001 (2020).
\bibitem{kitaev} A.~Y.~Kitaev, Russ.\ Math.\ Surv.\ {\bf 52}, 1191 (1997).
\bibitem{acin} A.~Acin, E.~Jane, and G.Vidal, Phys.\ Rev.\ A {\bf 64}, 050302(R) (2001).
\bibitem{DAriano} G.~Mauro D'Ariano, P.~Lo Presti, and Matteo~G.~A.~Paris, J.\ Opt.\ B {\bf 4},  273 (2002).
\bibitem{sacchi1} M.~F.~Sacchi, Phys.\ Rev.\ A {\bf 71}, 062340 (2005).
\bibitem{sacchi2} M.~F.~Sacchi, Phys.\ Rev.\ A {\bf 72}, 014305 (2005).
\bibitem{wang} G.~Wang and M.~Ying, Phys.\ Rev.\ A {\bf 73}, 042301 (2006).
\bibitem{pirandola} S.~Pirandola, B.~R.~Bardhan, T.~Gehring. C.~Weedbrook, and S.~Lloyd, Nat. Photonics {\bf 12}, 724 (2018).
\bibitem{hillery} M.~Hillery, H.~Gupta, and C.~Zhan, Phys.\ Rev.\ A {\bf 107}, 012435 (2023).
\bibitem{zhan} C.~Zhan, H.~Gupta, and M.~Hillery, ACM Transactions on Quantum Computing {\bf 5}, (2024) and arXiv 2306:17401 (2003).
\bibitem{ali} N.~Ali and M.~Hillery, Phys.\ Rev.\ A {\bf 110}, 012619 (2024).
\bibitem{Zhuang1} Quntao Zhuang and S.~Pirandola, Commun.\ Phys. {\bf 3},103 (2020).
\bibitem{Zhuang2} Quntao Zhuang and S.~Pirandola, Phys.\ Rev.\ Lett.\ {\bf 125}, 080505 (2020).
\bibitem{Pereira} J.~L.~Pereira, Quntao Zhuang and S.~Pirandola, Phys.\ Rev.\ Research {\bf 2}, 043189 (2020).
\bibitem{helstrom} C.~W.~Helstrom, \emph{Quantum Detection and Estimation Theory} (Academic, New York, 1976).
\bibitem{review} For reviews of state discrimination see Discrimination of Quantum States by J.~A.~Bergou, U.~Herzog, and M.~Hillery in \emph{Quantum State Estimation}, edited by M.~G.~A.~Paris and J.~\v{R}eha\v{c}ek (Springer Verlag, Berlin, 2004), and S.~M.~Barnett and S.~Croke, Advances in Optics and Photonics {\bf 1}, 238 (2009) and arXiv:0810.1970.
\bibitem{chuang1}  Z.~E.~Chin, D.~R.~Leibrandt, and I.~L.~Chuang, Phys.\ Rev.\ Lett.\  {\bf 134}, 210802 (2025).
\bibitem{chuang2} Z.~E.~Chin and I.~L.~Chuang, Phys.\ Rev.\ A {\bf 111}, 052447 (2025).
\bibitem{gupta} C.~Zhan and H.~Gupta, Proceedings of the IEEE International Conference on Quantum Computing and Engineering, 659 (2023).
\bibitem{ban} M.~Ban, K.~Kurokawa, R.~Momose, and O.~Hirota, Int.\ J.\ of Th.\ Phys.\ {\bf 36}, 1269 (1997).
\bibitem{eldar} Y.~C.~Eldar, A.~Megretski, and G.~C.~Verghese, IEEE Trans\ Inf.\ Theory {\bf 50}, 1198 (2004).
\bibitem{hausladen} P.~Hausladen and W.~Wootters, J.\ Mod.\ Opt.\ {\bf 41}, 2385 (1994).
\bibitem{eldar2} Y.~C.~Eldar and G.~D.~Forney, Jr., IEEE Trans\ Inf.\ Theory {\bf 47}, 858 (2001).
\bibitem{bagan} E.~Bagan, private communication.
\bibitem{barnett}S.~M.~Barnett, Phys.\ Rev.\ A {\bf 64}, 030303 (2001).


\end{thebibliography}
\end{document}